\documentclass[12pt,onecolumn,journal]{IEEEtran}

\usepackage{cite}
\usepackage[cmex10]{amsmath}
\usepackage{subfigure}
\ifCLASSINFOpdf
  \usepackage[pdftex]{graphicx}
\else
  \usepackage[dvips]{graphicx}
  \graphicspath{{../figs/}}
  \DeclareGraphicsExtensions{.eps}
\fi

\begin{document}
\title{GLRT-Based Spectrum Sensing with Blindly Learned Feature under Rank-1 Assumption}
\author{Peng~Zhang,~\IEEEmembership{Student Member,~IEEE,}
        Robert~Qiu,~\IEEEmembership{Senior Member,~IEEE,}
\thanks{The authors are with the Wireless Networking System Lab in Department of Electrical and Computer Engineering, Center for Manufacturing Research, Tennessee Technological University, Cookeville, TN, 38505, USA. E-mail: pzhang21@students.tntech.edu, \{rqiu, nguo\}@tntech.edu}}
\maketitle
\begin{abstract}
Prior knowledge can improve the performance of spectrum sensing. Instead of using universal features as prior knowledge, 
we propose to blindly learn the localized feature at the secondary user. Motivated by pattern recognition in machine learning, we define signal feature as the leading \textit{eigenvector} of the signal's sample covariance matrix. 
Feature learning algorithm (FLA) for blind feature learning and feature template matching algorithm (FTM) for spectrum sensing are proposed. Furthermore, we implement the FLA and FTM in hardware. Simulations and hardware experiments show that signal feature can be learned blindly. In addition, by using signal feature as prior knowledge, the detection performance can be improved by about 2 dB. 
Motivated by experimental results, we derive several GLRT based spectrum sensing algorithms under rank-1 assumption, considering signal feature, signal power and noise power as the available parameters. The performance of our proposed algorithms is tested on both synthesized rank-1 signal and captured DTV data, and compared to other state-of-the-art covariance matrix based spectrum sensing algorithms. In general, our GLRT based algorithms have better detection performance.
In addition, algorithms with signal feature as prior knowledge are about 2 dB better than algorithms without prior knowledge.

\end{abstract}
\IEEEpeerreviewmaketitle
\section{Introduction}
Radio frequency (RF) is fully allocated for primary users (PU), but it is not utilized efficiently \cite{staple2004end}. \cite{force2002spectrum, cabric2004implementation} show that the utilization of allocated spectrum only ranges from $15\%$ to $85\%$. This is even lower in rural areas. Cognitive radio (CR) is proposed so that secondary users (SU) can occupy the unused spectrum from PU, therefore improving the spectrum efficiency and enabling more RF applications. Spectrum sensing is the key function in CR. Each SU should be able to sense PU's existence accurately in low signal-to-noise ratio (SNR) to avoid interference.

Spectrum sensing can be casted as the signal detection problem. The detection performance depends on the available prior knowledge. If the signal is fixed and known to the receiver, matched filter gives the optimum detection performance \cite{SSSurvey2009,haykin2009spectrum,zeng2010review}. If signal is unknown, signal samples can be modeled as independent and identically distributed (i.i.d.) random variables, as well as noise. In such model, energy detector gives the optimum performance. However, though energy detector is blind to signal, it is not blind to noise. \cite{tandra2005fundamental} show that actual noise power is not obtainable and noise uncertainty problem can heavily limit energy detector's performance. In addition, signal is usually oversampled at the receiver, and non-white wide-sense stationary (WSS) model is more appropriate for signal samples.

\begin{figure}[tbp]
	\centering
		\includegraphics[width=0.50\textwidth]{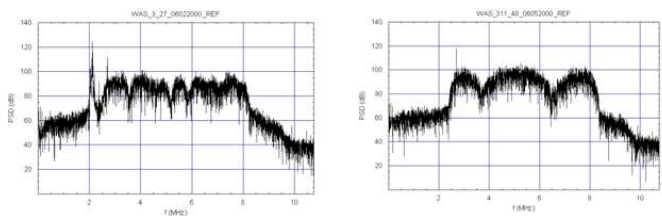}
	\caption{Spectrum measured at different locations in Washington D.C.. Left: `Single Family Home'; Right: `Apartment (High-Rise)'. The pilot tones are located at different frequency locations. Two spectrum suffer different frequency selective fading.}
	\label{fig:Two_Spectrum_WAS}
\end{figure}

Prior knowledge of PU signal is often considered in spectrum sensing algorithms. One class of spectrum sensing algorithms utilize prior knowledge from universal pre-determined signal spectral information. Take spectrum sensing algorithms for DTV signal for example. Pre-determined spectral information includes pilot tone \cite{cordeiro2007spectrum}, spectrum shape \cite{Quan2009ss} and cyclostationarity \cite{dandawate1994statistical}, etc. Generally speaking, they have good performance when it is assumed that those pre-determined features are universal. However, such assumption is not true in practice. From IEEE 802.22 DTV measurements \cite{DTV2006Measurements} as shown in Fig. \ref{fig:Two_Spectrum_WAS}, spectral features are location dependent due to different channel characteristics and synchronization mis-match, etc. Therefore, we cannot rely on universal pre-determined signal features for spectrum sensing. Furthermore, these non-blind algorithms are only limited to DTV signals.
In this paper, we propose to use localized signal feature learned at SU for spectrum sensing, so that feature can be location dependent. Motivated from pattern recognition in machine learning \cite{fukunaga1990introduction}, we define the signal feature as the leading \textit{eigenvector} of signal's sample covariance matrix. According to DKLT \cite{therrien1992discrete,watanabe1965kle}, there are two interesting properties:
\begin{enumerate}
	\item The leading \textit{eigenvector} is stable over time for non-white WSS signal while random for white noise.
	\item The leading \textit{eigenvector} for non-white WSS signal is most robust against white noise.
\end{enumerate}

Using these properties, we develop the feature learning algorithm (FLA) for blind feature learning and the feature template matching (FTM) algorithm using blindly learned signal feature for spectrum sensing. Noise uncertainty problem is avoided because actual noise power is not used in spectrum sensing. 

We blindly measure the feature similarity of the 25 seconds DTV data samples captured in Washington D.C. every 4.6 ms. Surprisingly, all features are almost exactly the same. 
In addition, simulation results will show that the detection performance of FTM can be improved by 2 dB, compared with other covariance matrix based algorithms without any prior knowledge. 

Motivated by the above simulation results, we have a patent disclosure for both algorithms \cite{patent2010} and verify them using Lyrtech hardware platform with FPGA and DSP \cite{secon2010,letter2010}. We have spent 2 man-months to implement FLA and FTM in Lyrtech hardware platform. 
We have performed a blind feature learning experiment in non-line-of-sight (NLOS) environment showing feature's stability over time. Moreover, we compare the detection performance of FTM and CAV in hardware as well. In the experiment, FTM is about 3 dB better than CAV without any prior knowledge.

We further develop various algorithms using the general likelihood ratio test (GLRT) method with signal feature as one of the available parameters. Unfortunately, close form results for GLRT are not always obtainable \cite{kay1998fundamentals}. For mathematical convenience, we derive GLRT algorithms under rank-1 assumption. There are three parameters for analysis: signal power, noise variance and signal feature. It is very convenient to perform analysis under rank-1 GLRT framework. We derive GLRT algorithms for 5 cases listed in Table \ref{GLRT_combinations}. 
\begin{table}
	\label{GLRT_combinations}
	\centering
	\caption{Cases Considering Available Parameters Under Rank-1 Assumption}
	\begin{tabular}{|c|c|c|c|}
		\hline
		Cases & Signal Power & Noise Power & Signal Feature \\
		\hline
		Case 1 & Yes & Yes & Yes\\
		\hline
		Case 2 & No & Yes & Yes\\
		\hline
		Case 3 & No & No & Yes\\
		\hline
		Case 4 & No & Yes & No\\
		\hline
		Case 5 & No & No & No\\
		\hline
	\end{tabular}
\end{table}
We use both rank-1 signal and captured DTV signal for simulation. Though DTV signal is not rank-1, simulation results have shown relative very good detection performance for our derived GLRT algorithms. Overall, among algorithms without noise uncertainty problem, our GLRT based algorithm in Case 3 and FTM with signal feature as prior knowledge is about 2 dB better than algorithms without prior knowledge. Interestingly, Case 3 with feature as prior knowledge is only slightly better than FTM, within 0.1 dB, though FTM has much lower computational complexity. In addition, our GLRT based algorithm in Case 5 is slightly better than AGM, which is the counterpart algorithm for Case 5 derived in \cite{lim2008glrt,SG_GLRT} without rank-1 assumption. 

Our algorithm derivations are different with those in \cite{lim2008glrt,SG_GLRT} in that:
\begin{enumerate}
	\item For the first time, we use the properties of the \textit{eigenvector}, not \textit{eigenvalue}, for spectrum sensing.
	\item For the first time, we analyze the problem with signal feature as one of the parameters using GLRT method.
	\item We derive GLRT based algorithms under rank-1 assumption.
\end{enumerate}

During the preparation of this paper, we notice that \cite{Wang2010GLRT} has also derived several GLRT based algorithms using the latest results from the \textsl{finite-sample optimality} of GLRT \cite{FiniteGLRT2009}. \cite{Wang2010GLRT} has shown interesting results by introducing prior distribution of unknown parameters to obtain better performance, compared with classical GLRT. We will extend our current work based on the methods in \cite{Wang2010GLRT,FiniteGLRT2009}.

The organization of this paper is as follows. Section \ref{System_Model} describes the system model. Section \ref{Blind_Feature_Learning} reviews the blind feature learning results from our previous work. We present our proposed GLRT based spectrum sensing algorithms in Section \ref{Detection_Algorithms}. Simulation results are shown in Section \ref{Simulation_Results} and conclusions are made in Section \ref{Conclusions}.

\section{System Model}
\label{System_Model}
We consider the case when there is one receive antenna to detect one PU signal. Let $r\left( t \right)$ be the continuous-time received signal at receiver after unknown channel with unknown flat fading. $r\left( t \right)$ is sampled with period $T_s$, and the received signal sample is $r\left[ n \right] = r\left( {nT_s } \right)$. In order to detect PU signal's existence, we have two hypothesis:
\begin{equation}
\begin{array}{l}
 {\cal H}_0 :r\left[ n \right] = w\left[ n \right] \\ 
 {\cal H}_1 :r\left[ n \right] = s\left[ n \right] + w\left[ n \right] \\ 
 \end{array}
\end{equation}
where $w\left[n\right]$ is the zero-mean white Gaussian noise, and $s\left[n\right]$ is the received PU signal after unknown channel and is zero-mean non-white WSS Gaussian. Two probabilities are of interest to evaluate detection performance: Detection probability, $P_d \left( {{\cal H}_1 |r\left[ n \right] = s\left[ n \right] + w\left[ n \right]} \right)$ and false alarm probability $P_f \left( {{\cal H}_1 |r\left[ n \right] = w\left[ n \right]} \right)$.

Assume spectrum sensing is performed upon the statistics of the $i^{th}$ sensing segment $\Gamma_{r,i}$ consisting of $N_s$ sensing vectors:
\begin{equation}
\Gamma_{r,i} = \left\{ {{{\bf{r}}_{(i-1)N_s+1}},{{\bf{r}}_{(i-1)N_s + 2}}, \cdots {{\bf{r}}_{(i-1)N_s + N_s}}} \right\}
\end{equation}
with
\begin{equation}
\label{vector_r}
{\bf r}_i  = \left[ {r\left[ i \right],r\left[ {i + 1} \right], \cdots ,r\left[ {i + N - 1} \right]} \right]^T
\end{equation}
where $\left(  \cdot  \right)^T $ denotes matrix transpose. ${\bf r}_i \sim {\cal N}(0, {\bf R}_r)$, and ${\bf R}_r$ can be approximated by sample covariance matrix ${\hat{\bf R}}_r$:
\begin{equation}
\label{sample_cov}
\hat{{\bf R}}_{r} = \frac{1}{{N_s }}\sum\limits_{i = 1}^{N_s } {{\bf r}_i {\bf r}_i ^T }
\end{equation}
We will use ${\bf R}_r$ instead of $\hat{{\bf R}}_r$ for convenience. The eigen-decomposition of ${\bf R} _r$ is:
\begin{equation}
{{\bf{R}}_r} = {\Phi _r}{\Lambda _r}\Phi _r^T = \sum\limits_{i = 1}^N {{\lambda _{r,i}}{\phi _{r,i}}\phi _{r,i}^T} 
\end{equation}
where
\begin{equation}
{\Phi _r} = \left[ {\begin{array}{*{20}{c}}
{{\phi _{r,1}}}&{{\phi _{r,2}}}& \cdots &{{\phi _{r,N}}}
\end{array}} \right]
\end{equation}
and
\begin{equation}
{\Lambda _r} = diag\left\{ {{\lambda _{r,1}},{\lambda _{r,2}}, \cdots ,{\lambda _{r,N}}} \right\}
\end{equation}
$diag\left\{ \cdot \right\}$ denotes the diagonal matrix, $\left\{ {\phi _{r,i} } \right\}$ are eigenvectors of ${\bf R}_r$ and $\left\{ {\lambda _{r,i} } \right\}$ are eigenvalues of ${\bf R}_r$, satisfying $\lambda _{r,1}  \ge \lambda _{r,2}  \ge ... \ge \lambda _{r,N}$. Accordingly, we have ${\bf R}_s$, $\Phi_s$ and $\Lambda_s$ for ${\bf s}_i$; ${\bf R}_w = \sigma^2 {\bf I}$ for ${\bf w}_i$, where ${\bf I}$ is identity matrix.

One practical issue is that noise $w\left[ n \right]$ after analog-to-digital converter (ADC) is usually non-white, due to RF characteristics. A noise whitening filter is commonly applied before ADC and the details can be found in \cite{zeng2007covariance}. In this paper, $r\left[n \right]$ can be viewed as received sample after the noise whitening filter. Therefore, $w \left[n \right]$ is white and $s \left[n \right]$ has taken noise whitening filter into account. In the rest of this paper, all noise is considered as white. 
\section{Blind Feature Learning}
\label{Blind_Feature_Learning}
In this section, we will briefly review our previous blind feature learning results. 

\begin{figure}
	\centering
		\includegraphics[width=0.50\textwidth]{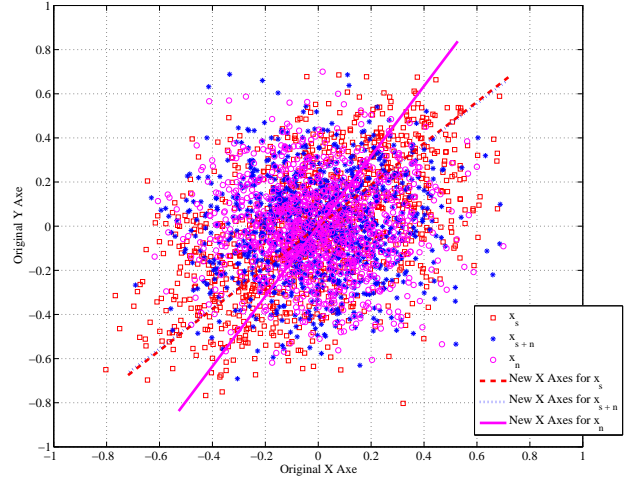}
	\caption{Feature of non-white signal is robust. Feature of white noise has randomness.}
	\label{fig:New_Directions_of_Signals}
\end{figure}

DKLT gives the optimum solution in searching signal subspace with maximum signal energy, which are represented by the eigenvectors \cite{therrien1992discrete,watanabe1965kle}. The leading eigenvector, a.k.a. feature, has maximum signal subspace energy, which is the leading eigenvalue. Moreover, feature is robust against noise and stable if the signal samples are non-white WSS. If signal samples are white, feature is random. This can be illustrated in the following way. Geometrically, feature is the new axes with largest projected signal energy \cite{therrien1992discrete,fukunaga1990introduction}. Let ${\bf x}_s$ be a $2 \times 1$ zero-mean non-white Gaussian random vector and ${\bf x}_n$ be a $2 \times 1$ zero-mean white Gaussian random vector. ${\bf x}_s$ and ${\bf x}_n$ have same energy and let ${\bf x}_{s+n} = {\bf x}_s + {\bf x}_n$. We plot 1000 samples of ${\bf x}_s$, ${\bf x}_n$ and ${\bf x}_{s + n}$ on a two dimensional graph in Fig. \ref{fig:New_Directions_of_Signals}. It can be seen that new X axes, a.k.a. feature, of ${\bf x}_s$ and ${\bf x}_{s+n}$ are exactly the same, while feature of ${\bf x}_n$ is rotated with some random degree.
We can use this property to differentiate non-white WSS signal $s\left[ n \right]$ and noise $w\left[ n\right]$. Let $N \times 1$ vector $\varphi_i$ be the extracted feature from the covariance matrix of sensing segment ${\Gamma}_{r,i}$. We obtain two consecutive features $\varphi_{i}$ and $\varphi_{i+1}$ from ${\Gamma}_{r,i}$ and ${\Gamma}_{r,i+1}$, respectively. If $\varphi_i$ and $\varphi_j$ are highly similar, then the signal feature is learned. We use the intuitive template matching method to define the similarity of $\varphi_i$ and $\varphi_j$: 
\begin{equation}
\label{similarity}
\rho _{i,j}  = \mathop {\max }\limits_{l = 1,2,...,N - k + 1} | {\sum\limits_{k = 1}^N {\varphi _i \left[ k \right]\varphi _j \left[ {k + l} \right]} } |
\end{equation}
The FLA is outlined as follows:
\begin{enumerate}
	\item Extract features $\varphi_i$ and $\varphi_{i+1}$ from two consecutive sensing segments $\Gamma_{r,i}$ and $\Gamma_{r,i + 1}$.
	\item Compute similarity $\rho _{i,i + 1}$ between these two features using (\ref{similarity}).
	\item If $\rho _{i,i + 1}  > T_e$ feature is learned as $\phi_{s,1} = \varphi _{i + 1}$, where $T_e$ is the threshold that can be determined empirically.
\end{enumerate}

We use captured DTV signal in Washington D.C. with 25 seconds duration \cite{DTV2006Measurements} to illustrate that feature is robust and stable over time. We measure feature similarity of consecutive sensing segments over the 25-second data with 4.6 ms per sensing segment. It is surprising that the feature extracted from the first sensing segment and the last sensing segment has similarity as high as $99.98\%$. Furthermore, signal feature is almost unchanged for about $99.46\%$ amount of time in 25 seconds. Signal feature is very robust and stable over time.

With learned signal feature $\phi_{s,1}$ as prior knowledge, we develop the intuitive FTM for spectrum sensing. FTM simply compare the similarity between the feature $\phi_{r,1}$ extracted from the new sensing segment $\Gamma_{r,i}$ and the signal feature $\phi_{s,1}$. If $\phi_{r,1}$ and $\phi_{s,1}$ are highly similar, PU signal exists. The FTM is outlined as follows:
\begin{enumerate}
	\item Extract feature $\phi_{r,i}$ from sensing segment $\Gamma_{r,i}$.
	\item ${\cal H}_1$ is true if:\\
\begin{equation}
\label{FTM_old}
{T_{FTM}} = \mathop {\max }\limits_{l = 1,2,...,N - k + 1} \left| {\sum\limits_{k = 1}^N {{\phi _{s,1}}\left[ k \right]{\phi_{r,1}}\left[ {k + l} \right]} } \right| > \gamma 
\end{equation}
where $\gamma$ is the threshold determined by desired $P_f$.
\end{enumerate}

Simulation results of FTM on DTV signal will be shown in Section \ref{Simulation_Results}. Approximately 2 dB gain will be obtained over algorithms without any prior knowledge. 

\begin{figure}[tbp]
	\centering
 		\includegraphics[width=0.50\textwidth]{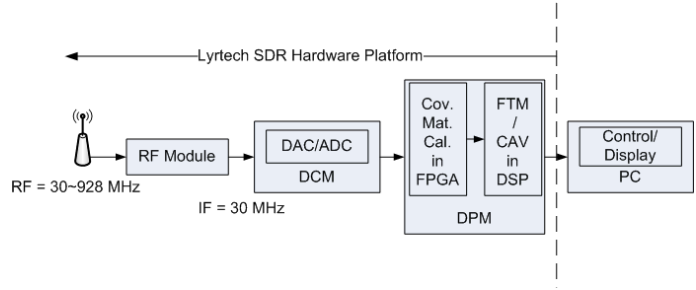}
	\caption{The top-level architecture of the spectrum sensing receiver. DCM: Digital conversion module. DPM: Digital processing module. }
	\label{fig:Arch}
\end{figure}

We have implemented the FLA, FTM in Lyrtech software-defined-radio (SDR) hardware platform. Another spectrum sensing algorithm based on sensing segment $\Gamma_{r,i}$, CAV \cite{zeng2007covariance}, is also implemented in the same hardware as well. CAV uses exactly the same signal as FTM, but CAV does not require any prior knowledge. It is considered as a blind benchmark algorithm for comparison purpose. The top-level architecture is illustrated in Fig. \ref{fig:Arch}. Covariance matrix calculation is implemented in Xilinx Virtex 4 SX35 FPGA; feature extraction (leading eigenvector calculation), similarity calculation and CAV are implemented in TI C64x+ DSP. Leading eigenvector calculation is the major challenge in our implementation. Since FLA and FTM only use the leading eigenvector for feature learning and spectrum sensing, we use FPCA \cite{frieze2004fast} and the computational complexity is reduced from ${\cal O} (N^3)$ to ${\cal O} (N^2)$. Without much effort in implementation optimization, the leading eigenvector can be extracted within 20 ms. 
We perform the blind feature learning experiment in an NLOS indoor environment. PU signal is emulated by sinusoidal generated from Rohde \& Schwarz signal generator. Transmit antenna and receive antenna are 2 meters away, and the direct patch is blocked by the signal generator. A $-50$ dBm sinusoidal signal at $435$ MHz is transmitted. SU's RF is tuned to $432$ MHz center frequency with $20$ MHz bandwidth. Channel, signal frequency, signal power and noise power are unknown to the receiver. In the experiment, our hardware platform record the feature similarities of consecutive sensing segments around every 20 ms for 20 seconds with $N_s = 2^{20}$ and $N = 32$. By setting $T_e = 80\%$, $\rho_{i,i+1} > T_e$ for $87.6\%$ amount of time when PU signal exists. The similarity of features extracted from the first segment and the last segment is $94.3\%$. As a result, signal feature in this experiment is very stable and robust over time.

\begin{figure}[tbp]
	\centering
		\includegraphics[width=0.45\textwidth]{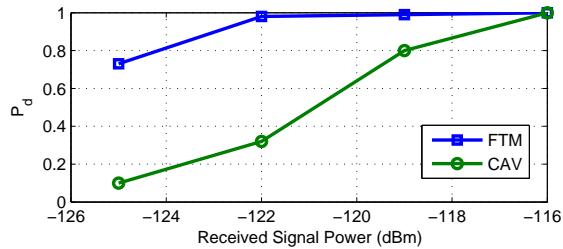}
		\caption{$P_d$ VS Received Signal Power at $P_f = 10\%$ for FTM and CAV.}
	\label{fig:Detection_Hardware}
\end{figure}

Then, we set the feature extracted from the last sensing segment as learned signal feature $\phi_{s,1}$ and perform the spectrum sensing experiment. FTM is compared with CAV \cite{zeng2007covariance}, which is totally blind. In order to compare the detection performance of both algorithms under the same SNR, we connect the signal generator to the receiver with SMA cable. PU feature is already stored as $\phi_{s,1}$ at the receiver. We vary the transmit power of the signal generator from $-125$ dBm to $-116$ dBm with $3$ dB increments. Cable loss is omitted and transmit signal power is considered as received signal power. $1000$ measurements are made for each setting. The $P_d$ VS Received Signal Power curves at $P_f = 10\%$ for FTM and CAV are depicted in Fig. \ref{fig:Detection_Hardware}. It can be seen that to reach $P_d \approx 100\%$, the required minimum received signal power for CAV is at least 3 dB more than FTM. 
The above hardware experiments show that feature is very robust and stable over time, and feature can be learned blindly. We can use feature as prior knowledge to obtain better spectrum sensing performance.
\section{Detection Algorithms}
\label{Detection_Algorithms}
%

We try to derive GLRT based algorithms considering feature as one of the available parameters. In this paper's GLRT based algorithms, signal is detectable \textsl{iff.} $\lambda_{s,1} > \sigma^2$. Signal detection under low energy coherence (LEC) condition ($\lambda_{s,1} < \sigma^2$) \cite{vantrees1971} is not considered. All algorithms requiring $\sigma^2$ as prior knowledge have noise uncertainty problem, because the actual $\sigma^2$ is not obtainable \cite{tandra2005fundamental}.
\subsection{GLRT Based Detection Algorithms}
\subsubsection{Background Review}
Since $s\left[n\right]$ and $w\left[n\right]$ are uncorrelated, the distribution of received signal vector ${\bf r}_i$ under two hypothesis can be represented as:
\begin{equation}
{\cal H}_0 : {\bf r}_i \sim {\cal N}(0, \sigma^2{\bf I})
\end{equation}
and
\begin{equation}
{\cal H}_1: {\bf r}_i \sim {\cal N}(0, {\bf R}_s + \sigma^2{\bf I})
\end{equation}
The detection will be based upon the statistics of $N_s$ sensing vectors in $\Gamma_{r,i}$, say, $\Gamma_{r,1}$. If $\left\{ {\bf r}_i \right\}$ are i.i.d., we have:
\begin{equation}
\label{iid_assumption}
\left\{ \begin{array}{l}
p\left( {{\Gamma _{r,1}}|{\cal H}_0} \right) = \prod\limits_{i = 1}^{{N_s}} {p\left( {{{\bf{r}}_i}|{\cal H}_0} \right)} \\
p\left( {{\Gamma _{r,1}}|{\cal H}_1} \right) = \prod\limits_{i = 1}^{{N_s}} {p\left( {{{\bf{r}}_i}|{\cal H}_1} \right)} 
\end{array} \right.
\end{equation}

Though $\left\{{\bf r}_i\right\}$ defined in (\ref{vector_r}) are not i.i.d., we will use (\ref{iid_assumption}) for mathematical convenience. The likelihood function for $\Gamma_{r,1}$ under ${\cal H}_0$ condition can be:
\begin{equation}
\begin{aligned}
	p\left( {\Gamma_{r,1}|{\cal H}_0} \right) & = \prod\limits_{i = 1}^{{N_s}} {p\left( {{{\bf{r}}_i}|{\cal H}_0} \right)} \\
&	= \prod\limits_{i = 1}^{{N_s}} {\frac{1}{{{{\left( {2\pi {\sigma ^2}} \right)}^{N/2}}}}\exp \left[ { - \frac{1}{{2{\sigma ^2}}}{\bf{r}}_i^T{\bf{r}}_i} \right]} 
\end{aligned}
\end{equation}
and the corresponding logarithm likelihood function is:
\begin{equation}
\label{lnp_H0}
\ln p\left( {\Gamma_{r,1}|{\cal H}_0} \right) = - \frac{{N{N_s}}}{2}\ln \left( {2\pi {\sigma ^2}} \right) - \frac{1}{{2{\sigma ^2}}}\sum\limits_{i = 1}^{{N_s}} {{\bf{r}}_i^T{{\bf{r}}_i}}
\end{equation}

The logarithm likelihood function under ${\cal H}_1$ is:
\begin{equation}
\begin{aligned}
p\left( {\Gamma_{r,1}|{\cal H}_1} \right) & = \prod\limits_{i = 1}^{{N_s}} {p\left( {{{\bf{r}}_i}|{\cal H}_1} \right)} \\
& = \prod\limits_{i = 1}^{{N_s}} {\frac{1}{{{{\left( {2\pi } \right)}^{\frac{N}{2}}}{{\det }^{\frac{1}{2}}}\left( {{{\bf{R}}_s} + {\sigma ^2}{\bf{I}}} \right)}}\exp \left[ { - \frac{1}{2}{\bf{r}}_i^T{{\left( {{{\bf{R}}_s} + {\sigma ^2}{\bf{I}}} \right)}^{ - 1}}{\bf{r}}_i} \right]}
\end{aligned}
\end{equation}
and the corresponding logarithm likelihood function is:
\begin{equation}
\label{lnp_H1}
\ln p\left( {\Gamma_{r,1}|{\cal H}_1} \right) = - \frac{{N{N_s}}}{2}\ln 2\pi 
- \frac{1}{2}\left[ {{N_s}\sum\limits_{i = 1}^N {\ln \left( {{\lambda _{s,i}} + {\sigma ^2}} \right)}  + \sum\limits_{j = 1}^{{N_s}} {\sum\limits_{i = 1}^N {\frac{{{{\left( {\phi _{s,i}^T{{\bf{r}}_j}} \right)}^2}}}{{{\lambda _{s,i}} + {\sigma ^2}}}} } } \right]
\end{equation}

In signal detection, it is desired to design an algorithm maximizing the $P_d$ for a given $P_f$. According to Neyman-Pearson theorem, this can be done by the likelihood ratio test (LRT):
\begin{equation}
\label{LRT}
L\left( {\Gamma_{r,1}} \right) = \frac{{p\left( {\Gamma_{r,1}|{\cal H}_{1} } \right)}}{{p\left( {\Gamma_{r,1}|{\cal H}_0 } \right)}}
\end{equation}
or
\begin{equation}
\label{LLT}
\ln L\left( \Gamma_{r,1} \right) = \ln p\left( {\Gamma_{r,1}|{\cal H}_1 } \right) - \ln p\left( {\Gamma_{r,1}|{\cal H}_0 } \right)
\end{equation}

${\cal H}_1$ is true if $L\left( \Gamma_{r,1} \right)$ or $\ln L \left( \Gamma_{r,1} \right)$ is greater than a threshold $\gamma$, which is determined by desired $P_f$.

In practice, however, it is usually not possible to know the exact likelihood functions. If one or several parameters are unknown, composite hypothesis testing is used. GLRT is a common method in composite hypothesis testing problems. GLRT first gets a maximum likelihood estimate (MLE) of the unknown parameters set ${\bf \Theta }$ under ${\cal H}_0$ and ${\cal H}_1$:
\begin{equation}
\begin{array}{l}
 {\bf \hat \Theta }_0  = \mathop {\arg \max }\limits_{{\bf \Theta }_0 } p\left( {\Gamma_{r,1}|{\bf \Theta }_0, {\cal H}_0 } \right) \\ 
 {\bf \hat \Theta }_1  = \mathop {\arg \max }\limits_{{\bf \Theta }_1 } p\left( {\Gamma_{r,1}|{\bf \Theta }_1, {\cal H}_1 } \right) \\ 
 \end{array}
\end{equation}
where ${\bf \Theta }_0 $ and ${\bf \Theta }_1$ are the unknown parameters under ${\cal H}_0$ and ${\cal H}_1$, respectively. ${\cal H}_1$ is true if:
\begin{equation}
\label{L_G}
L_G \left( \Gamma_{r,1} \right) = \frac{{p\left( {\Gamma_{r,1}|{\bf \hat \Theta }_1 ,{\cal H}_1 } \right)}}{{p\left( {\Gamma_{r,1}|{\bf \hat \Theta }_0 ,{\cal H}_0 } \right)}} > \gamma 
\end{equation}
or
\begin{equation}
\ln {L_G}\left( \Gamma_{r,1} \right) = 
\ln p\left( {\Gamma_{r,1}|{{{\bf{\hat \Theta }}}_1},{\cal H}_1} \right) - \ln p\left( {\Gamma_{r,1}|{{{\bf{\hat \Theta }}}_0},{\cal H}_0} \right) > \gamma 	\label{ln_L_G}
\end{equation}

Unfortunately, sometimes closed-form solutions for GLRT cannot be derived directly \cite{kay1998fundamentals}. For mathematical convenience, we will assume the signal covariance matrix to be rank-1 matrix. According to DKLT \cite{therrien1992discrete}, the optimum rank-1 approximated matrix for ${\bf R}_s$ is 
\begin{equation}
{\bf R}_s^1 = \lambda_{s,1} \phi_{s,1}\phi_{s,1}^T
\end{equation}
There are three parameters available under rank-1 assumption: $\lambda_{s,1}$, $\sigma^2$ and $\phi_{s,1}$. Notice that signal feature $\phi_{s,1}$ is also one of the parameters. Therefore, it is very convenient to analyze our feature based spectrum sensing algorithm under the rank-1 GLRT framework. We list the algorithms correspondent to different combinations of available parameters in Table \ref{GLRT_combinations}. Case 1 is for upper benchmark reference assuming all parameters known. Except for Case 1, we do not consider $\lambda_{s,1}$ as prior knowledge, because it is impractical to assume the signal energy of PU as prior knowledge.

Under rank-1 assumption, only $\lambda_{s,1} \neq 0$ and (\ref{lnp_H1}) becomes:
\begin{equation}
\label{lnp_H1_new}
\begin{aligned}
&	\ln p\left( {{\Gamma _{r,1}}|{\cal H}_1} \right) = \\
& - \frac{{N{N_s}}}{2}\ln 2\pi 
- \frac{{{N_s}}}{2}\left[ {\ln \left( {{\lambda _{s,1}} + {\sigma ^2}} \right) + \sum\limits_{j = 1}^{{N_s}} {\frac{{{{\left( {\phi _{s,1}^T{{\bf{r}}_j}} \right)}^2}}}{{{N_s}\left( {{\lambda _{s,1}} + {\sigma ^2}} \right)}}} } \right] 
- \frac{{{N_s}}}{2}\left[ {\left( {N - 1} \right)\ln \left( {{\sigma ^2}} \right) + \sum\limits_{j = 1}^{{N_s}} {\sum\limits_{i = 2}^N {\frac{{{{\left( {\phi _{s,i}^T{{\bf{r}}_j}} \right)}^2}}}{{{N_s}{\sigma ^2}}}} } } \right]
\end{aligned}
\end{equation}

Since $\Phi _s \Phi _s^T  = \sum\limits_{i = 1}^N {\phi _{s,i} \phi _{s,i}^T }  = {\bf I}$, we have:
\begin{equation}
\label{relation_phi}
\sum\limits_{i = 2}^N {\phi _{s,i} \phi _{s,i}^T }  = {\bf I} - \phi _{s,1} \phi _{s,1}^T
\end{equation}

With (\ref{relation_phi}), $\sum\limits_{i = 2}^N {{{\left( {\phi _{s,i}^T{{\bf{r}}_j}} \right)}^2}}$ in (\ref{lnp_H1_new}) becomes:
\begin{equation} 
\begin{aligned} 
  \sum\limits_{i = 2}^N {{{\left( {\phi _{s,i}^T{{\bf{r}}_j}} \right)}^2}} & = {\bf{r}}_j^T\left( {\sum\limits_{i = 2}^N {{\phi _{s,i}}\phi _{s,i}^T} } \right){{\bf{r}}_j}  \\
	&   = {\bf{r}}_j^T\left( {{\bf{I}} - {\phi _{s,1}}\phi _{s,1}^T} \right){{\bf{r}}_j} \\
	&   = {\bf{r}}_j^T{{\bf{r}}_j} - {\left( {\phi _{s,1}^T{{\bf{r}}_j}} \right)^2} \label{sub_phi}
\end{aligned}
\end{equation}

Notice that:
\begin{equation}
\label{Cov_R}
\frac{1}{{{N_s}}}\sum\limits_{j = 1}^{{N_s}} {{{\bf{r}}_j}{\bf{r}}_j^T}  = {{\bf{R}}_r}
\end{equation}
\begin{equation} 
	\begin{aligned}
		\frac{1}{N_s}\sum\limits_{j = 1}^{N_s} {\bf r}_j{\bf r}_j^T & = trace\left( {\bf R}_r \right)\\
		& = \sum\limits_{i = 1}^N \lambda _{r,i}	\label{Cov_R_3}
	\end{aligned}
\end{equation}
\begin{equation}
\begin{aligned}
	\frac{1}{{{N_s}}}\sum\limits_{j = 1}^{{N_s}} {{{\left( {\phi _{s,1}^T{{\bf{r}}_j}} \right)}^2}} & = \frac{1}{{{N_s}}}\phi _{s,1}^T\sum\limits_{j = 1}^{{N_s}} {{\bf{r}}_j^T{{\bf{r}}_j}} {\phi _{s,1}} \\
	& = \phi _{s,1}^T{{\bf{R}}_r}{\phi _{s,1}} \label{Cov_R_2}
\end{aligned}
\end{equation}

Together with (\ref{Cov_R_3}) and (\ref{lnp_H0}), we have:
\begin{equation}
\label{lnp_H0_simp}
\ln p\left( {{\Gamma _{r,1}}|{\cal H}_0} \right) =  - \frac{{{N_s}}}{2}\left[ {N\ln \left( {2\pi {\sigma ^2}} \right) + \frac{1}{{{\sigma ^2}}}\sum\limits_{i = 1}^N {{\lambda _{r,i}}} } \right]
\end{equation}

Together with (\ref{Cov_R_3}), (\ref{Cov_R_2}) and (\ref{lnp_H1_new}), we have:
\begin{equation}
\label{lnp_H1_simp}
\begin{aligned}
&	\ln p\left( {{\Gamma _{r,1}}|{\cal H}_1} \right) = \\
&	- \frac{{N{N_s}}}{2}\ln 2\pi - \frac{{{N_s}}}{2}\left[ {\ln \left( {{\lambda _{s,1}} + {\sigma ^2}} \right) + \frac{{\phi _{s,1}^T{{\bf{R}}_r}{\phi _{s,1}}}}{{{\lambda _{s,1}} + {\sigma ^2}}}} \right]
- \frac{{{N_s}}}{2}\left[ {\left( {N - 1} \right)\ln \left( {{\sigma ^2}} \right) + \frac{{\left( {\sum\limits_{i = 1}^N {{\lambda _{r,i}} - \phi _{s,1}^T{{\bf{R}}_r}{\phi _{s,1}}} } \right)}}{{{\sigma ^2}}}} \right]
\end{aligned}
\end{equation}

We will use (\ref{lnp_H0_simp}) and (\ref{lnp_H1_simp}) extensively to derive GLRT based algorithm considering 5 cases in Table \ref{GLRT_combinations}.
\subsubsection{Case 1: All parameters available}
In this case, we have the classical estimator-correlator (EC) test \cite{kay1998fundamentals}. ${\cal H}_1$ is true if:
\begin{equation}
\begin{aligned}
	{T_{EC}} & = \sum\limits_{j = 1}^{{N_s}} {{\bf{r}}_j^T{{\bf{R}}_s}{{\left( {{{\bf{R}}_s} + {\sigma ^2}{\bf{I}}} \right)}^{ - 1}}{{\bf{r}}_j}} \\
 & = \frac{{{N_s}}}{N}\sum\limits_{i = 1}^N {\frac{{{\lambda _{s,i}}}}{{{\lambda _{s,i}} + {\sigma ^2}}}\phi _{s,i}^T{{\bf{R}}_r}{\phi _{s,i}}}  > \gamma	\label{T_EC}
\end{aligned}
\end{equation}
Details of this derivation can be found in \cite{kay1998fundamentals}, using eigen-decomposition properties. 

Under rank-1 assumption, we can get the new test by replacing ${\bf R}_s$ with ${\bf R}_s^1$ in (\ref{T_EC}) and only $\lambda_{s,1} \neq 0$. By ignoring corresponding constants, ${\cal H}_1$ is true if:
\begin{equation}
\label{case_1_new}
{T_{CASE1}} = \frac{{{\lambda _{s,1}}}}{{{\lambda _{s,1}} + {\sigma ^2}}}\phi _{s,1}^T{{\bf{R}}_r}{\phi _{s,1}} > \gamma
\end{equation}
\subsubsection{Case 2: $\sigma^2$ and $\phi_{s,1}$ available}
In this case, we need to get MLE of $\lambda_{s,1}$. By taking the derivative to (\ref{lnp_H1_simp}) with respect to $\lambda_{s,1}$, we have: 
\begin{equation} 
\frac{{\partial \ln p\left( {{\Gamma _{r,1}}|{\lambda _{s,1}},{\cal H}_1} \right)}}{{\partial {\lambda _{s,1}}}} = 
- \frac{{{N_s}}}{2}\left[ {\frac{1}{{{\lambda _{s,1}} + {\sigma ^2}}} - \frac{{\phi _{s,1}^T{{\bf{R}}_r}{\phi _{s,1}}}}{{{{\left( {{\lambda _{s,1}} + {\sigma ^2}} \right)}^2}}}} \right] \label{lnp_lambda_c2}
\end{equation}

Let $\frac{{\partial \ln p\left( {{\Gamma _{r,1}}|{\lambda _{s,1}},{\cal H}_1} \right)}}{{\partial {\lambda _{s,1}}}} = 0$ and we have MLE of $\lambda_{s,1}$:
\begin{equation}
\label{lambda_C2_H1}
{{\hat \lambda }_{s,1}} = \phi _{s,1}^T{{\bf{R}}_r}{\phi _{s,1}} - {\sigma ^2}
\end{equation}

Together with (\ref{ln_L_G}), (\ref{lnp_H0_simp}), (\ref{lnp_H1_simp}) and (\ref{lambda_C2_H1}) and ignoring the constants, we can get the test for Case 2. ${\cal H}_1$ is true if:

\begin{equation}
\label{case_2_new}
{T_{CASE2}} = \phi _{s,1}^T{{\bf{R}}_r}{\phi _{s,1}} > \gamma
\end{equation}
where $\gamma$ depends on the noise variance $\sigma^2$.
\subsubsection{Case 3: $\phi_{s,1}$ available}
This is the case when only signal feature is known. We need to get MLE of $\lambda_{s,1}$ and $\sigma^2$. By taking the derivative to (\ref{lnp_H0_simp}) with respect to $\sigma^2$, we have: 
\begin{equation}
\label{sigma_H0}
\frac{{\partial \ln p\left( {{\Gamma _{r,1}}|{\sigma ^2},{\cal H}_0} \right)}}{{\partial {\sigma ^2}}} =  - \frac{{{N_s}}}{2}\left[ {\frac{N}{{{\sigma ^2}}} - \frac{{\sum\limits_{i = 1}^N {{\lambda _{r,i}}} }}{{{{\left( {{\sigma ^2}} \right)}^2}}}} \right]
\end{equation}

Let $\frac{{\partial \ln p\left( {\Gamma_{r,1}|{\sigma ^2},{\cal H}_0} \right)}}{{\partial {\sigma ^2}}} = 0$ and we have the MLE of $\sigma^2$ under ${\cal H}_0$:
\begin{equation}
\label{sigma_0}
{\hat {\sigma} }_0^2 = \sum\limits_{i = 1}^N {{\lambda _{r,i}}} /N
\end{equation}

By taking the derivative to (\ref{lnp_H1_simp}) with respect to $\lambda_{s,1}$, we have:
\begin{equation} 
\frac{{\partial \ln p\left( {{\Gamma _{r,1}}|{\lambda _{s,1}},{\sigma ^2},{\cal H}_1} \right)}}{{\partial {\lambda _{s,1}}}} = 
- \frac{{{N_s}}}{2}\left( {\frac{1}{{{\lambda _{s,1}} + {\sigma ^2}}} - \frac{{\phi _{s,1}^T{{\bf{R}}_r}{\phi _{s,1}}}}{{{{\left( {{\lambda _{s,1}} + {\sigma ^2}} \right)}^2}}}} \right)	\label{lnp_lambda_c3}
\end{equation}

Let $ \frac{{\partial \ln p\left( {\Gamma_{r,1}|{\lambda_{s,1}}, \sigma^2, {{\cal H}_1}} \right)}}{{\partial {\lambda _{s,1}}}} = 0$ and we have
\begin{equation}
\label{lambda_C3_H1}
{{\hat \lambda }_{s,1}} + \hat \sigma _1^2 = \phi _{s,1}^T{{\bf{R}}_r}{\phi _{s,1}}
\end{equation}

Then, by taking the derivative to (\ref{lnp_H1_simp}) with respect to $\sigma^2$, we have:
\begin{equation}
\label{rank1_H1_der}
\begin{aligned}
&	\frac{{\partial \ln p\left( {\Gamma_{r,1}|\lambda_{s,1},\sigma^2,{\cal H}_1} \right)}}{{\partial {\sigma ^2}}} =	\\
&	- \frac{N_s}{2}\left[ {\frac{1}{{\lambda _{s,1}  + \sigma ^2 }} - \frac{{\phi _{s,1}^T{{\bf{R}}_r}{\phi _{s,1}}}}{{\left( {\lambda _{s,1}  + \sigma ^2 } \right)^2 }}} \right]
- \frac{N_s}{2}\left[ {\frac{{N - 1}}{{\sigma ^2 }} - \frac{1}{{\left( {\sigma ^2 } \right)^2 }}\left( {{\sum\limits_{i = 1}^N {{\lambda _{r,i}}} } - \left( {\phi _{s,1}^T{{\bf{R}}_r}{\phi _{s,1}}} \right)^2 } \right)} \right]
\end{aligned}
\end{equation}

Let $ \frac{{\partial \ln p\left( {\Gamma_{r,1}|{\lambda_{s,1}},{\sigma^2},{{\cal H}_1}} \right)}}{{\partial {\sigma ^2}}} = 0$ and together with (\ref{lambda_C3_H1}), we have
\begin{equation}
\label{sigma_C3_H1}
{\hat {\sigma} } _1^2 = \left( {\sum\limits_{i = 1}^N {{\lambda _{r,i}}}  - \phi _{s,1}^T{{\bf{R}}_r}{\phi _{s,1}}} \right)/\left( {N - 1} \right)
\end{equation}

$\hat \sigma _1^2$ can be interpreted as the average energy mapped onto the non-signal subspace. 

Together with (\ref{sigma_0}), (\ref{sigma_C3_H1}), (\ref{lambda_C3_H1}) and (\ref{ln_L_G}), we can get the test for Case 3. Therefore, ${\cal H}_1$ is true if:
\begin{equation}
\label{case_3_new}
{T_{CASE3}} = \ln \frac{{{ {\hat {\sigma} }}_0^2}}{{\phi _{s,1}^T{{\bf{R}}_r}{\phi _{s,1}}}} + \left( {N - 1} \right)\ln \frac{{{ {\hat {\sigma} }}_0^2}}{{{ {\hat {\sigma} }}_1^2}} > \gamma
\end{equation}
where $\hat \sigma_0^2$ and $\hat \sigma_1^2$ are represented in (\ref{sigma_0}) and (\ref{sigma_C3_H1}).
\subsubsection{Case 4: $\sigma^2$ available}
In this case, we need to get MLE of $\lambda_{s,1}$ and $\phi_{s,1}$. The logarithm of the likelihood function under ${\cal H}_0$ is (\ref{lnp_H0_simp}), which can be used directly for the likelihood ratio test. 

By taking the derivative to (\ref{lnp_H1_simp}) with respect to $\lambda_{s,1}$, we have similar result but with known $\sigma^2$ and the estimate of $\phi_{s,1}$:
\begin{equation}
\label{lnp_lambda_c4}
\hat \lambda _{s,1}  + \sigma ^2  = {\hat \phi _{s,1}^T{{\bf{R}}_r}{{\hat \phi }_{s,1}}}
\end{equation}

MLE finds $\phi_{s,1}$ that maximize $\ln p\left( {\Gamma_{r,1}|\phi _{s,1} ,H_1 } \right)$ in (\ref{lnp_H1_simp}). (\ref{lnp_H1_simp}) can be rewritten as:
\begin{equation}
\label{case4_lnp_H1}
\ln p\left( {\Gamma_{r,1}|{\lambda_{s,1}},{\phi_{s,1}},{\cal H}_1} \right) = - \frac{N_s}{2}\left( {\frac{1}{{{\lambda _{s,1}} + {\sigma ^2}}} - \frac{1}{{{\sigma ^2}}}} \right){{\phi _{s,1}^T{{\bf{R}}_r}{\phi _{s,1}}}}
+ g\left( {{\sigma ^2},{\lambda _{s,1}}} \right)
\end{equation}
where $g\left( {{\sigma ^2},{\lambda _{s,1}}} \right)$ is the function including all other terms in (\ref{lnp_H1_simp}).

Since $ - \frac{1}{2}\left( {\frac{1}{{{\lambda _{s,1}} + {\sigma ^2}}} - \frac{1}{{{\sigma ^2}}}} \right) > 0$, $\ln p\left( {\Gamma_{r,1}|{\lambda_{s,1}},{\phi_{s,1}},{\cal H}_1} \right)$ is monotonically increasing with regard to ${\phi _{s,1}^T{{\bf{R}}_r}{\phi _{s,1}}}$. The MLE of $\phi_{s,1}$ is the solution to the following optimization problem:
\begin{equation}
\label{problem_phi}
\begin{array} {c}
  \mathop {\arg \max }\limits_{{\phi _{s,1}}} \;\;\;\;{{\phi _{s,1}^T{{\bf{R}}_r}{\phi _{s,1}}}} \\
  s.t. \; \; \; \; \; \phi _{s,1}^T \phi _{s,1}  = 1
\end{array}
\end{equation}

The solution can be found by Lagrange multipliers method. Let
\begin{equation}
\label{l_func}
f\left( {{\phi _{s,1}},\alpha } \right) = {{\phi _{s,1}^T{{\bf{R}}_r}{\phi _{s,1}}}} + \alpha \left( {\phi _{s,1}^T{\phi _{s,1}} - 1} \right)
\end{equation}
Let the derivative to $f\left( {{\phi _{s,1}}} \right)$ with respect to $\phi_{s,1}$ and $\alpha$ be zero respectively:
\begin{equation}
\label{phi_est}
\begin{array}{l}
{\bf R}_r{\phi _{s,1}} = \alpha {\phi _{s,1}}\\
\phi _{s,1}^T{\phi _{s,1}} = 1
\end{array}
\end{equation}
Therefore, ${{\hat \phi }_{s,1}}$ is the leading eigenvector of ${\bf R}_r$ and $\alpha$ is the leading eigenvalue of ${\bf R}_r$. The MLE of $\phi_{s,1}$:
\begin{equation}
\label{phi_C4_H1}
{{\hat \phi }_{s,1}} = \phi _{r,1}
\end{equation}

With (\ref{sigma_0}), (\ref{lnp_H1_simp}), (\ref{lnp_lambda_c4}), (\ref{phi_C4_H1}) and (\ref{ln_L_G}), we have
\begin{equation}
\label{case_4}
{T_{case4}} = \frac{{{\lambda _{r,1}}}}{{{\sigma ^2}}} - \ln \frac{{{\lambda _{r,1}}}}{{{\sigma ^2}}} - 1
\end{equation}

Since function $f\left( x \right) = x - \ln x - 1$ is monotonically increasing with regard to $x$, ${\cal H}_1$ is true if:
\begin{equation}
\label{case_4_new}
{T_{CASE4}} = {\lambda _{r,1}} > \gamma
\end{equation}
where $\gamma$ depends on the noise variance $\sigma^2$. 

It is interesting that (\ref{case_4_new}) is essentially the same as the signal-subspace eigenvalues (SSE) in \cite{lim2008glrt} under rank-1 assumption. Ignoring the constant terms in SSE, ${\cal H}_1$ is true if:
\begin{equation}
T_{SSE}  = \frac{{\sum\limits_{i = 1}^{N'} {\lambda _{r,i} } }}{{\sigma ^2 }} - \ln \frac{{\prod\limits_{i = 1}^{N'} {\lambda _{r,i} } }}{{\sigma ^2 }} > \gamma 
\end{equation}
where $N'$ corresponds to the largest $i$ such that $\lambda_{r,i} > \sigma^2$. If signal is rank-1, $N'$ can be 0 or 1. If $N' = 0$, $\lambda_{r,1} < \sigma^2$. If $N' = 1$, SSE becomes:
\begin{equation}
\label{SSE1}
T_{SSE1}  = \frac{{\lambda _{r,1} }}{{\sigma ^2 }} - \ln \frac{{\lambda _{r,1} }}{{\sigma ^2 }}
\end{equation}

Since (\ref{SSE1}) is monotonically increasing with regard to $\lambda_{r,1}$ and $\sigma^2$ is constant, the test can be further simplified as:
\begin{equation}
T_{SSE1}  = \lambda _{r,1} 
\end{equation}

As a result, no matter $N' = 0$ or $1$, the test statistic will be $\lambda_{r,1}$, which is the same as (\ref{case_4_new}).
\subsubsection{Case 5: All parameters unavailable}
In this case, we need to get MLE of $\lambda_{s,1}$, $\sigma^2$ and $\phi_{s,1}$. By taking the derivative to (\ref{lnp_H0_simp}) with respect to $\sigma^2$, we have MLE of $\sigma^2$ under ${\cal H}_0$ in (\ref{sigma_0}). Using similar techniques in Case 3 and Case 4, we have the following MLE of $\lambda_{s,1}$, $\sigma^2$ and $\phi_{s,1}$:
\begin{equation}
\begin{array}{l}
{ {\hat \sigma}} _0^2 = \sum\limits_{i = 1}^N {{\lambda _{r,i}}} /N,\\
{ {\hat \sigma} }_1^2 = \sum\limits_{i = 2}^N {{\lambda _{r,i}}} /\left( {N - 1} \right),\\
{{{\hat \lambda} }_{s,1}} = {\lambda _{r,1}} - \sum\limits_{i = 2}^N {{\lambda _{r,i}}} /\left( {N - 1} \right),\\
{{{\hat \phi }}_{s,1}} = {\phi _{r,1}}
\end{array}
\end{equation}
and ${\cal H}_1$ is true if:
\begin{equation}
\label{case_5_new}
{T_{CASE5}} = \ln \frac{{{\bar {\hat {\sigma} }}_0^2}}{{{\lambda _{r,1}}}} + \left( {N - 1} \right)\ln \frac{{{\bar {\hat {\sigma} }}_0^2}}{{{\bar {\hat {\sigma} }}_1^2}} > \gamma
\end{equation}
\subsection{Covariance Matrix Based Algorithms}
Sample covariance matrix based spectrum sensing algorithms have been proposed. MME \cite{zeng2007maximum} and CAV \cite{zeng2007covariance} have no prior knowledge, while FTM \cite{Dyspan2011} has feature as prior knowledge. Another interesting algorithm is AGM \cite{lim2008glrt,SG_GLRT}, which is derived using (\ref{iid_assumption}) without considering the rank of ${\bf R}_s$ and prior knowledge. We call these algorithms covariance based because the first step of all these algorithms is to calculate the sample covariance matrix ${\bf R}_r$ from $\Gamma_{r,i}$. 
\subsubsection{MME}
MME is also derived using (\ref{iid_assumption}). ${\cal H}_1$ is true if:
\begin{equation}
\label{MME}
{T_{MME}} = \frac{{{\lambda _{r,1}}}}{{{\lambda _{r,N}}}} > \gamma 
\end{equation}
\subsubsection{CAV}
${\cal H}_1$ is true if:
\begin{equation}
\label{CAV}
{T_{CAV}} = \frac{{\sum\limits_{i = 1}^N {\sum\limits_{j = 1}^N {\left| {{r_{ij}}} \right|} } }}{{\sum\limits_{i = 1}^N {\left| {{r_{ii}}} \right|} }} > \gamma 
\end{equation}
where $r_{ij}$ are the elements of ${\bf R}_r$.
\subsubsection{FTM}
FTM has been introduced in (\ref{FTM_old}).
\subsubsection{AGM}
AGM is derived without considering the rank of original signal. ${\cal H}_1$ is true if:
\begin{equation}
\label{AGM}
T_{AGM}  = \frac{{\frac{1}{N}\sum\limits_{i = 1}^N {\lambda _{r,i} } }}{{\left( {\prod\limits_{i = 1}^N {\lambda _{r,i} } } \right)^{\frac{1}{N}} }} > \gamma 
\end{equation}

Among all algorithms without noise uncertainty problem, CAV do not need any eigen-decomposition at all and has lowest computational complexity. FTM only needs to calculate $\phi_{r,1}$ and this can be done using fast principal component analysis (F-PCA) \cite{sharma2007fast} with computational complexity ${\cal O} \left( N^2 \right)$. To the best of our knowledge, only CAV and FTM have been implemented and demonstrated in hardware platforms successfully.
\section{Simulation Results}
\label{Simulation_Results}
All algorithms to be simulated are summarized in Table II. EC uses the original ${\bf R}_s$, Case 1 -- Case 5 uses the algorithms under rank-1 assumption. Both Case 3 and FTM have the signal feature as prior knowledge. Case 5, MME, CAV and AGM have no prior knowledge. 
\begin{table}
	\label{Simulation_Algorithms}
	\centering
	\caption{Summary of the Algorithms for Simulation}
	\begin{tabular}{|c|c|c|c|}
		\hline
		Name & Test Statistics & Equation & Prior Knowledge\\
		\hline
		EC & $T_{EC}$ & (\ref{T_EC}) & ${\bf R}_s$, $\sigma^2$	\\
		\hline
		Case 1 & $T_{CASE1}$ & (\ref{case_1_new}) & $\lambda_{s,1}$, $\sigma^2$, $\phi_{s,1}$	\\
		\hline
		Case 2 & $T_{CASE2}$ & (\ref{case_2_new}) & $\sigma^2$, $\phi_{s,1}$	\\
		\hline
		Case 3 & $T_{CASE3}$ & (\ref{case_3_new}) & $\phi_{s,1}$	\\
		\hline
		Case 4 & $T_{CASE4}$ & (\ref{case_4_new}) & $\sigma^2$	\\
		\hline
		Case 5 & $T_{CASE5}$ & (\ref{case_5_new}) & None\\
		\hline
		MME & $T_{MME}$ & (\ref{MME}) & None \\
		\hline
		CAV & $T_{CAV}$ & (\ref{CAV}) & None \\
		\hline
		FTM & $T_{FTM}$ & (\ref{FTM_old}) & $\phi_{s,1}$ \\
		\hline
		AGM & $T_{AGM}$ & (\ref{AGM}) & None \\
		\hline
	\end{tabular}
\end{table}
Note that EC, Case 1, Case 2 and Case 4 have noise uncertainty problem, because the tests depends on the actual $\sigma^2$. Case 3, Case 5, MME, CAV, FTM and AGM, however, do not have noise uncertainty problem, because their tests do not depend on the actual $\sigma^2$. 
For each simulation, zero-mean i.i.d. Gaussian noise is added according to different SNR. 1000 simulations are performed on each SNR level and all algorithms are applied on the same noisy samples for each simulation. 

\subsection{Simulation with Rank-1 Signal}
\begin{figure}
	\centering
		\includegraphics[width=0.50\textwidth]{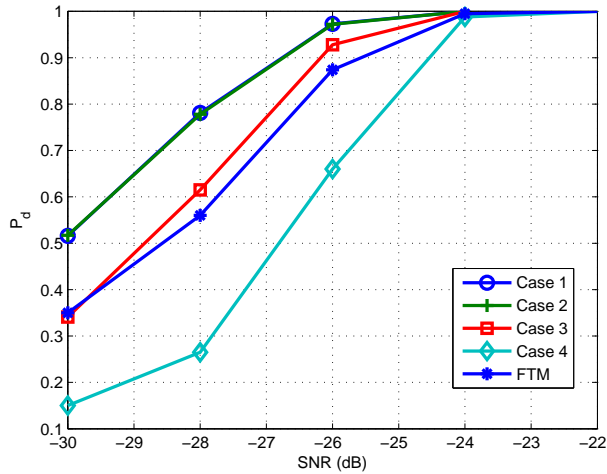}
	\caption{Algorithms with prior knowledge. $P_d$ at various SNR levels with $P_f = 10\%$, using rank-1 signal.}
	\label{fig:Detection_rank1_SNR_All}
\end{figure}

\begin{figure}
	\centering
		\includegraphics[width=0.50\textwidth]{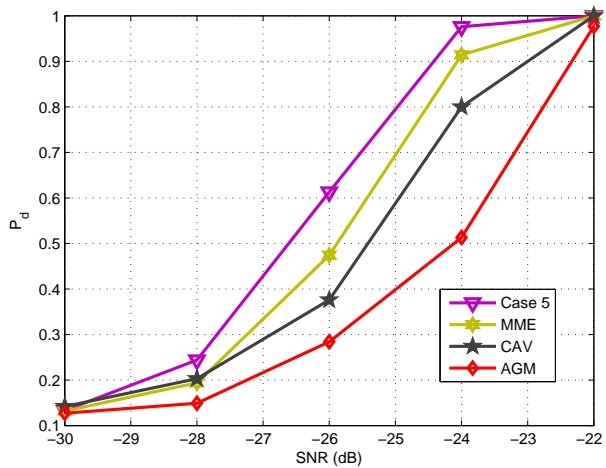}
	\caption{Algorithms without prior knowledge. $P_d$ at various SNR levels with $P_f = 10\%$, using rank-1 signal.}
	\label{fig:Detection_rank1_SNR_Blind}
\end{figure}

We first use simulated WSS rank-1 signal samples to perform Monte Carlo simulation. We use $N_s = 10^5$ samples to obtain rank-1 ${\bf R}_s$ with $N = 32$. Signal feature $\phi_{s,1}$ is obtained from ${\bf R}_s$. Since ${\bf R}_s$ is rank-1 matrix, EC is equivalent to Case 1. Fig. \ref{fig:Detection_rank1_SNR_All} shows the $P_d$ VS SNR plot with $P_f = 10\%$ for algorithms with prior knowledge while Fig. \ref{fig:Detection_rank1_SNR_Blind} shows the $P_d$ VS SNR plot with $P_f = 10\%$ for algorithms without prior knowledge.
From the simulation results, we can see that our derived GLRT based algorithms under rank-1 assumption work very well. To reach $P_d \approx 100\%$, EC requies about -24 dB SNR. It can be seen that Case 2 has almost the same performance with Case 1. This is because ${\lambda _{s,1}}/\left( {{\lambda _{s,1}} + {\sigma ^2}} \right)$ in (\ref{case_1_new}) is constant if $\sigma^2$ is stable and true to the detector, a.k.a., no noise uncertainty problem. As a result, (\ref{case_1_new}) and (\ref{case_2_new}) are using the same statistics and they are essentially equivalent. Case 3 with feature as prior knowledge is about 2 dB better than Case 4 with $\sigma^2$ as prior knowledge. Interestingly, the intuitive FTM is only slight worse than Case 3, though computational complexity for FTM is much lower than that of Case 3. Case 5 is slightly worse than Case 4, within 0.1 dB. Case 5 is about 1 dB better than MME, and 1.5 dB better than CAV. AGM, however, does not have comparable performance with other algorithms for rank-1 signal when SNR is low. 

Overall, among all algorithms without noise uncertainty problem, Case 3 and FTM with feature as prior knowledge are about 2 dB better than other algorithms when no prior knowledge available. Our derived GLRT based algorithm in Case 5 has best performance among all algorithms without prior knowledge.

\subsection{Simulation with Captured DTV Signal}

Now we use one sensing segment of DTV signal captured in Washington D.C. with $N_s = 10^5$ and $N = 32$ to test all algorithms.

\begin{figure}
	\centering
		\includegraphics[width=0.50\textwidth]{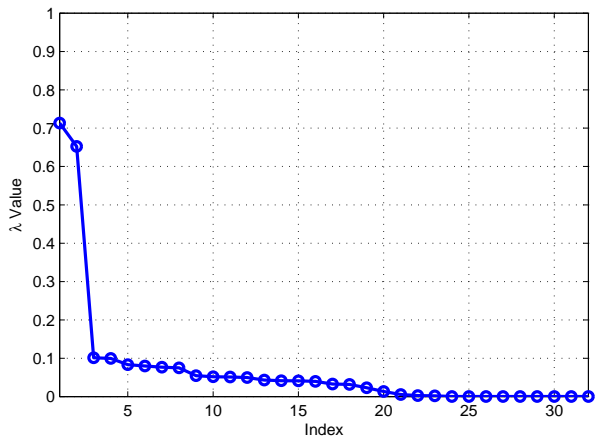}
	\caption{Normalized eigenvalue distribution of captured DTV signal. $N_s = 10^5$ and $N = 32$.}
	\label{fig:DTV_Eigen_Distr}
\end{figure}

\begin{figure}
	\centering
		\includegraphics[width=0.50\textwidth]{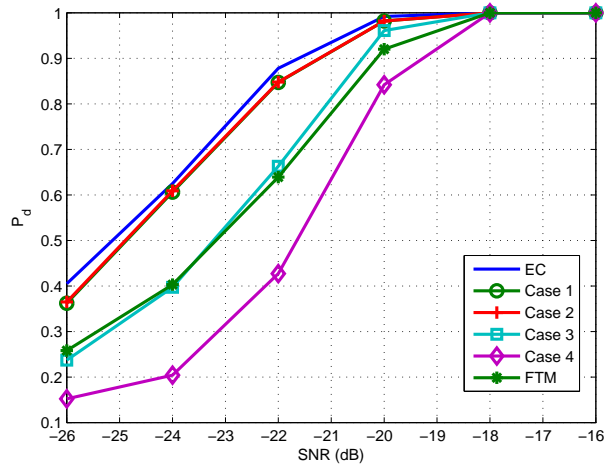}
	\caption{Algorithms with prior knowledge. $P_d$ at various SNR levels with $P_f = 10\%$, using captured DTV signal.}
	\label{fig:Detection_DTV_SNR_All}
\end{figure}

\begin{figure}
	\centering
		\includegraphics[width=0.50\textwidth]{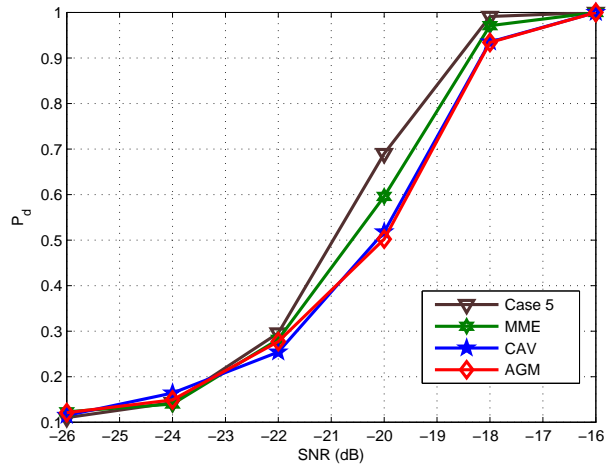}
	\caption{Algorithms without prior knowledge. $P_d$ at various SNR levels with $P_f = 10\%$, using captured DTV signal.}
	\label{fig:Detection_DTV_SNR_Blind}
\end{figure}

We first examine the rank of the signal. The normalized eigenvalue distribution of $R_s$ is plotted in Fig. \ref{fig:DTV_Eigen_Distr}. It is obvious that the rank of $R_s$ is greater than 1.
Then, we perform the Monte Carlo simulation to test the detection performance of all algorithms. 
Simulation results are shown in Fig. \ref{fig:Detection_DTV_SNR_All} for algorithms with prior knowledge while Fig. \ref{fig:Detection_DTV_SNR_Blind} shows the results for algorithms without prior knowledge. Both figures use $P_d$ VS SNR plot with $P_f = 10\%$. We can see that for DTV signal, all algorithms do not work as good as they are for the rank-1 signal. To reach $P_d \approx 100\%$, EC requires about -20 dB SNR. It can be seen that Case 1 using ${\bf R}_s^1$ is about 0.1 dB worse than EC using original ${\bf R}_s$. Again, Case 2 has the same performance with Case 1, because (\ref{case_1_new}) and (\ref{case_2_new}) are using the same statistics and they are essentially equivalent. Case 3 with feature as prior knowledge is about 2 dB better than Case 4 with $\sigma^2$ as prior knowledge. FTM has almost the same performance with Case 3. Case 4 is about 1 dB better than Case 5, MME, CAV and AGM, which are all blind. It can be seen that for non-rank-1 signal, AGM has almost the same performance as CAV. At -20 dB SNR, $P_d \approx 70\%$ for Case 5 while only $60\%$ and $52\%$ for MME and CAV/AGM, respectively. At -24 dB SNR, however, CAV and AGM have slightly higher $P_d$. 

Generally speaking, among all algorithms without noise uncertainty problem, Case 3 and FTM with feature as prior knowledge are 2 dB better than algorithms without prior knowledge. Among all algorithms without prior knowledge, our GLRT based algorithm in Case 5 is slightly better than MME, CAV and AGM. 
\section{Conclusions}
\label{Conclusions}
In this paper we considered the spectrum sensing for single PU with single antenna. Received signal is oversampled with unknown oversampling rate and modeled as a non-white WSS Gaussian process. Using the concept of pattern recognition in machine learning, we defined the signal feature as the leading eigenvector of the signal's sample covariance matrix. Our previous work has found that signal feature is robust against noise and stable over time. Both simulation and hardware experiments showed that signal feature can be learned blindly. In addition, by using signal feature as prior knowledge, the detection performance can be improved. 

Under rank-1 assumption of the signal covariance matrix, we derived several GLRT based algorithms for signal samples considering signal feature as one of the available parameters, as well as signal power and noise power. 

Rank-1 signal and captured DTV data were simulated with our derived GLRT based spectrum sensing algorithms and other state-of-the-art algorithms, including MME, CAV, FTM and AGM. MME, CAV and AGM can be viewed as the benchmark algorithms when no prior knowledge is available, while FTM can be viewed as the benchmark algorithm when only signal feature is available. The simulation results showed that our derived GLRT based algorithms have relatively better performance than the benchmark algorithms under the same available prior knowledge conditions. In general, algorithms with signal feature as prior knowledge are about 2 dB better than the algorithms without prior knowledge, and 2 dB worse than EC when all parameters are prior knowledge. Interestingly, the detection performance of FTM was almost the same as that of our GLRT based algorithm with signal feature as prior knowledge, though FTM has much lower computational complexity and has already been implemented in our previous work.

More generalized results under rank-k assumption will be discussed. New methods in \cite{Wang2010GLRT,FiniteGLRT2009} will be applied in our framework. Spectrum sensing for multiple antennas and cooperative spectrum sensing will also be discussed. Moreover, we will explore more machine learning techniques for cognitive radio, including robust principal component analysis \cite{candes2009robust}, fast low-rank approximations \cite{frieze2004fast}, manifold learning \cite{weinberger2006unsupervised}, etc. 

\section*{Acknowledgment}
The authors would like to thank Shujie Hou for helpful discussions. This work is funded by National Science Foundation, through two grants (ECCS-0901420 and ECCS-0821658), and Office of Naval Research, through two grants (N00010-10-1-0810 and N00014-11-1-0006).


\bibliographystyle{ieeetr}
\bibliography{bib/CR,bib/Pattern_Recognition} 

\ifCLASSOPTIONcaptionsoff
  \newpage
\fi

\end{document}